\documentclass[a4paper,fleqn,usenatbib]{mnras}

\usepackage{newtxtext,newtxmath}

\usepackage[T1]{fontenc}
\usepackage{ae,aecompl}

\usepackage{txfonts}
\usepackage{natbib}

\usepackage{graphicx}	
\usepackage{amsmath}	
\usepackage{amssymb}	

\title[ ]{On the redshift of the very high energy $\gamma$-ray BL Lac object 
 \\ S2 0109+22}

\author[S. Paiano et al.]{
S. Paiano,$^{1,2}$\thanks{E-mail: simona.paiano@oapd.inaf.it}
M. Landoni,$^{3}$
R. Falomo,$^{1}$
R.Scarpa,$^{4}$
and A. Treves,$^{5}$
\\
$^{1}$INAF, Osservatorio Astronomico di Padova, Vicolo dell'Osservatorio 5 I-35122 Padova (PD) - ITALY\\
$^{2}$INFN, Via Marzolo 8, I-35131 Padova - ITALY\\
$^{3}$INAF, Osservatorio Astronomico di Brera, Via E. Bianchi 46 I-23807 Merate (LC) - ITALY\\  
$^{4}$Instituto de Astrofisica de Canarias, C/O Via Lactea, s/n E38205 - La Laguna (Tenerife) - ESPANA\\
$^{5}$ Universit\`a degli Studi dell'Insubria, Via Valleggio 11 I-22100 Como - ITALY
}

\date{Accepted XXX. Received YYY; in original form ZZZ}

\pubyear{2016}

\begin{document}
\label{firstpage}
\pagerange{\pageref{firstpage}--\pageref{lastpage}}
\maketitle

\begin{abstract}
The determination of the redshift of extragalactic gamma ray sources is of fundamental importance for the interpretation of their emission models and extragalactic population studies, AGN classification schemes, and to study the interaction with the extragalactic background light. Here we focus on the BL Lac S2 0109+22, recently detected at very high energies. We obtained a high signal-to-noise optical spectrum at Gran Telescopio Canarias for the source over the spectral range 4000 - 10000 \AA, to search for spectral features of the host galaxy and/or of the nuclear source and it was compared with previous observations in order to confirm and/or dispute its redshift determination.
We find the S2 0109+22 optical spectrum is well represented by a power law continuum without any absorption or emission feature of Equivalent Width > 0.1 $\AA$. Based on the new optical spectrum we are able to set a redshift lower limit $z>$0.35, assuming the source is hosted by a massive elliptical galaxy typical for this class of sources.
The redshift $z$= 0.265 proposed in the literature for this object is most probably referred to a r=18.3 galaxy at 15'' from S2 0109. This galaxy could belong to a group of faint galaxies located at z$\sim$ 0.26.

\end{abstract}



\section{Introduction}

Blazars, and in particular the sub-class of the BL Lac objects \citep{falomo2014}, are radio-loud active galactic nuclei (AGNs) characterized by a non-thermal, variable and double-bump shape continuum flux spanning from radio to $\gamma$-ray band. Although they are the most numerous extragalactic population in the high energy (HE) and very high energy (VHE; E>100 GeV) bands, most of them have unknown or very uncertain redshift. The recognition of the redshift is a key factor to determine the luminosity of these sources, to characterize and build realistic emission models \citep[e.g.][]{fossati1998, bottcher2013} and to study the absorption of their high energy gamma-rays emitted with the extragalactic background light (EBL) \citep[see e.g.][]{franceschini2008, dominguez2011}. 

These considerations motivated several spectroscopic campaigns in order to measure the BL Lac redshift. 
We are undertaking spectroscopic campaign at the Gran Telescopio Canarias (GTC) to obtain high optical signal-to-noise ratio (SNR) spectra of TeV blazar with unknown or uncertain redshift. First results of this program were presented in \cite{landoni2015bis} for B0218+357 and \cite{landoni2015} for S2 0954+65.

In this paper we report on spectroscopic observations of the bright (V$\sim$15.5) BL Lac object S2 0109+22 that was recently detected in the VHE $\gamma$-ray band by the MAGIC Cherenkov telescopes \citep{mirzoyan2015}.
The new data allow us to clarify the contrasting literature results about the redshift of this source. 

In this work we assume the following cosmological parameters: H$_0=$ 70 km s$^{-1}$ Mpc$^{-1}$, $\Omega_{\Lambda}$=0.7, and $\Omega_{m}$=0.3.

\section{S2 0109+22}
The source S2 0109+22  (GC 0109+22, TXS 0109+224, RGB J0112+227) is a compact radio object \citep{davis1971, pauliny1972} with a 5GHz flux of 0.78 Jy.
The optical identification of the radio source was done with a stellar object of magnitude 15.5 through the optical Palomar Sky Survey \citep{Owen1977}.
The first spectroscopic observation of S2 0109+22, performed by \cite{wills1976}, did not find absorption or emission lines, suggesting a BL Lac nature for the object. Subsequent higher quality spectroscopic data\footnote{S2 0109+22 spectra are available in digital form in the ZBLLAC database (http://archive.oapd.inaf.it/zbllac/)} \citep{falomo1994, sbarufatti2006, sbarufatti2009, landoni2013} confirmed the featureless spectrum.
In contrast with the previous results, the optical spectrum obtained by \cite{healey2008} showed prominent galactic absorption lines at $z$=0.265. Curiously another spectrum obtained by \cite{shaw2009}, showed a pure featureless spectrum, but in \cite{shaw2012} galactic absorption features at $z$=0.265 were again present. 

\begin{figure}
   \centering
    \includegraphics[width=8.5cm]{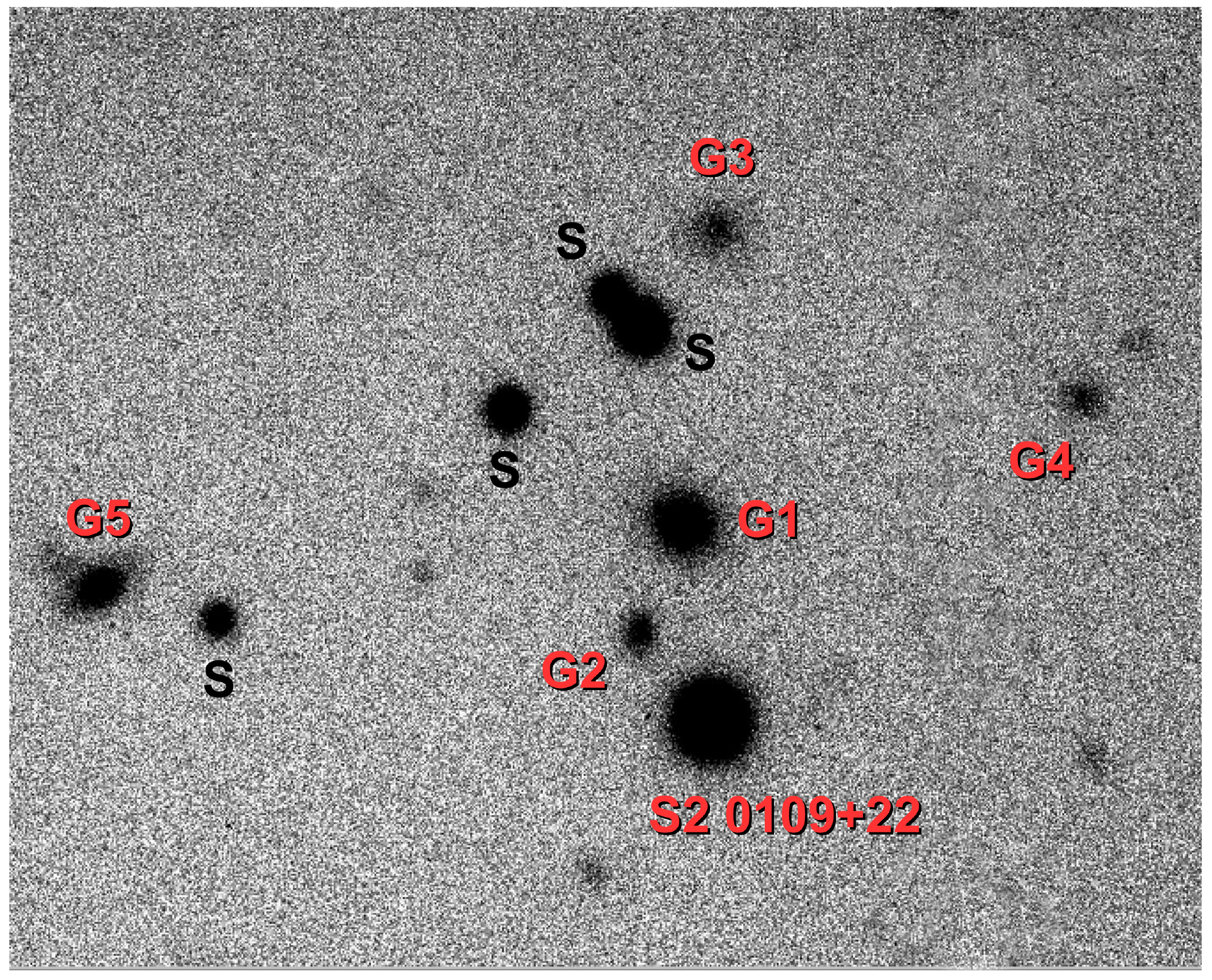}
      \caption{R-band optical image (field= 45 arcsec, North up, East left) obtained at the New Technology Telescope (NTT) for the BL Lac object S2 0109+22 \citep{falomo1996}. Five galaxies, labelled as G, are present in the field of view of the BL Lac object. The sources flagged as S are classified as star by SDSS.}
         \label{fig:NTT_image_R_BL0109}
   \end{figure}

From an optical image obtained during a low state (R mag=16.5) of the source, S2 0109+22 appears unresolved (\cite{falomo1996}, see also Fig. \ref{fig:NTT_image_R_BL0109}). Assuming the nucleus hosted by an elliptical galaxy with M$_R$=-23.5, a lower limit of the redshift of $z>$0.4 was proposed.
Following R-band images, obtained by \cite{nilsson2003}, also found the object unresolved.

At the higher frequencies, S2 0109+22 was originally revealed in X-ray bands by \textit{Einstein}, EXOSAT, ROSAT and \textit{Swift}, and by \textit{Fermi} in the gamma-ray band \citep[$F_{100MeV-100GeV}$= 7.78$\times$10$^{-11}$ erg cm$^{-2}$ s$^{-1}$, ][]{acero2015}.
Finally, the source was also detected in July 2015 at VHE (E>100 GeV)  gamma rays by the MAGIC telescopes \citep{mirzoyan2015} during a high-flux state monitored by \textit{Fermi}.

From multiwavelength observations, S2 0109+22 is classified as Intermediate BL Lac (IBL) object by \cite{padovani1995} and \cite{ghisellini2011} based on the $\alpha_{ro}$ - $\alpha_{oX}$ diagram and the position of the synchrotron peak frequency. The strong optical and radio flux variability and SED properties are consistent with this classification \citep{ciprini2003, ciprini2004}.

\section{Observations and data analysis}
S2 0109+22 was observed on 18 September 2015 with the 10.4 m GTC telescope at the Roque de Los Muchachos, in La Palma. 
The observations were performed with the spectrograph OSIRIS \citep{cepa2003} using two grisms, R1000B and R1000R, in order to cover the whole spectral range  4000-10000 $\AA$, and with a slit width = 1'' yielding a spectral resolution $\lambda$/$\Delta\lambda$ = 800.  For each grism we obtained three individual exposures of 250 sec in order to perform optimal cleaning of cosmic rays. 
During the observations the source was of 14.8 in R band as derived by a short exposure acquisition image.

\begin{figure*}
   \centering
   \includegraphics[width=10cm, height=15cm, angle=-90]{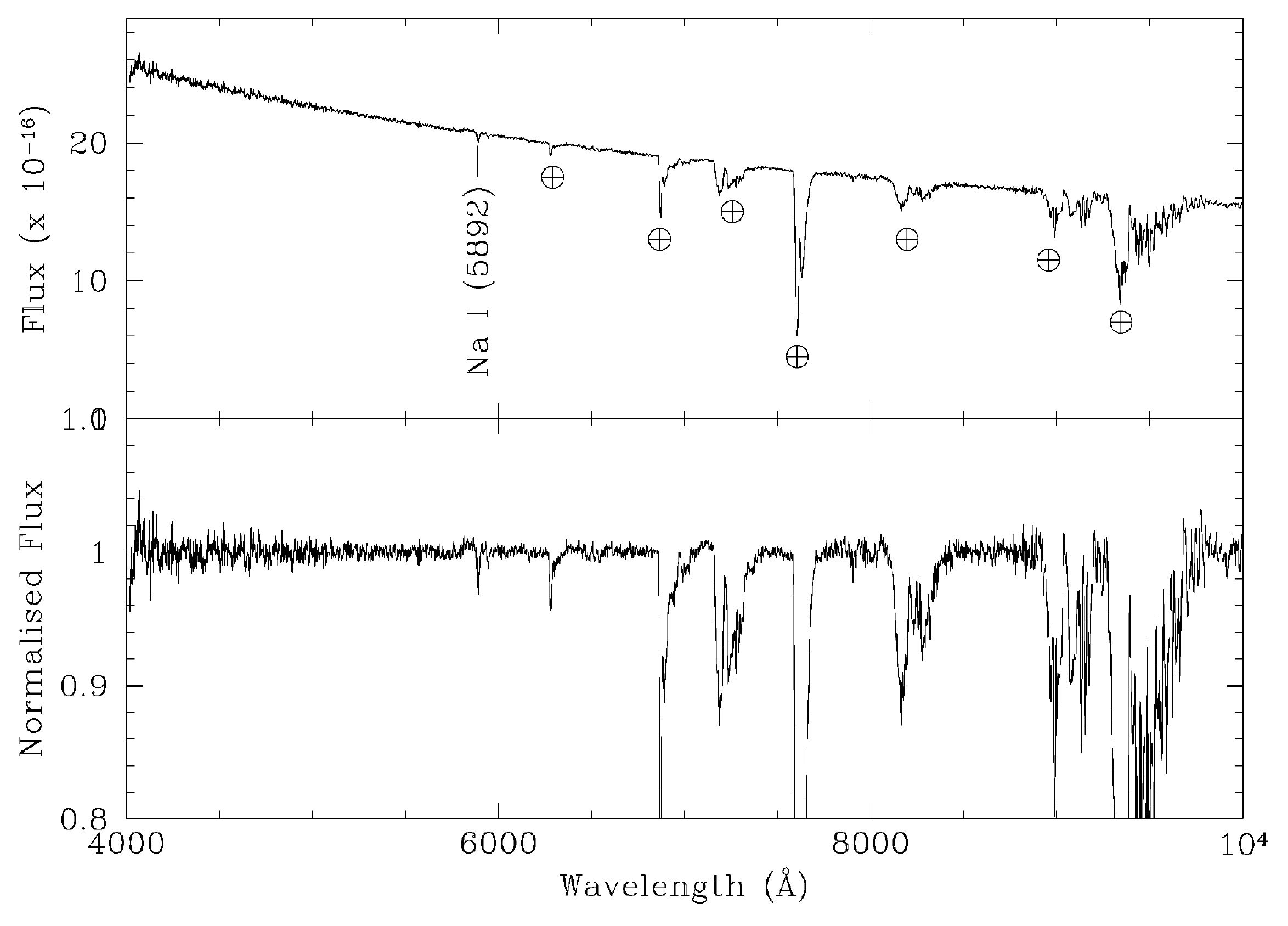}
   \caption{The optical spectrum of the BL Lac object S2 0109+22 obtained with OSIRIS@GTC. \textit{Upper panel}: Flux calibrated spectrum well described by a power law with a spectral index $\alpha$=-0.56$\pm$0.06. Telluric absorption O$_2$ and H$_2$0 bands are marked by $\oplus$. The only visible interstellar absorption line is Na I (5892$\AA$). \textit{Bottom panel}: Normalised spectrum.}
              \label{fig:S0109_overall_flux}%
    \end{figure*}

Data reduction was carried out using IRAF\footnote{IRAF (Image Reduction and Analysis Facility) is distributed by the National Optical Astronomy Observatories, which are operated by the Association of Universities for Research in Astronomy, Inc., under cooperative agreement with the National Science Foundation.} and standard procedures for long slit spectroscopy.
The accuracy of the wavelength calibration is 0.1 $\AA$.
Flux calibration was obtained from the observations of a spectro-photometric standard star during the same night.
The exposures of the two grisms were combined and a final spectrum covering the whole spectral range was merged (see Fig. \ref{fig:S0109_overall_flux}). 
The SNR per pixel of this spectrum ranges from minimum $\sim$150 in the blue region to 400 in the central band. 
The continuum is very well fitted by a power law $F(\lambda)\propto \lambda^{\alpha}$ with a spectral index $\alpha$=-0.56$\pm$0.06.

\section{Results and Conclusions}
\begin{figure*}
   \centering
   \includegraphics[width=8.0cm, angle=-90]{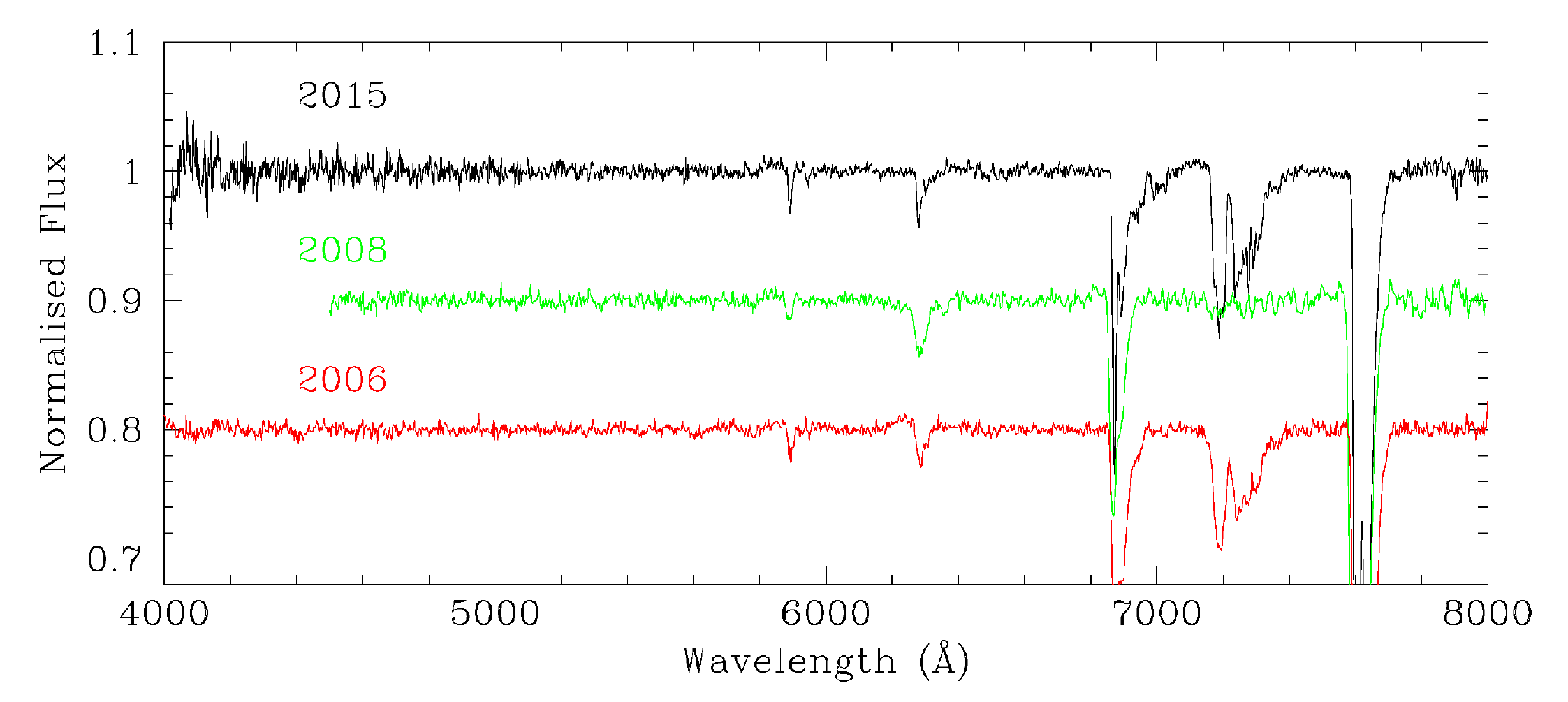}
   \caption{Comparison of the normalised spectra of S2 0109+22 taken in different epochs. \textit{From top to bottom}: 2015 spectrum (SNR$\sim$650); 2008 spectrum (SNR$\sim$500) \citep{landoni2013}; 2006 spectrum (SNR$\sim$500) by \citep{sbarufatti2009}. SNR estimated at 5500$\AA$ for resolution element of 15$\AA$. The 2008 and 2006 spectra are shifted of 0.1 and 0.2 respectively. }
    \label{fig:S0109_superimp_spectra}%
    \end{figure*}

\begin{figure}
   \centering
   \includegraphics[width=6cm, angle=-90]{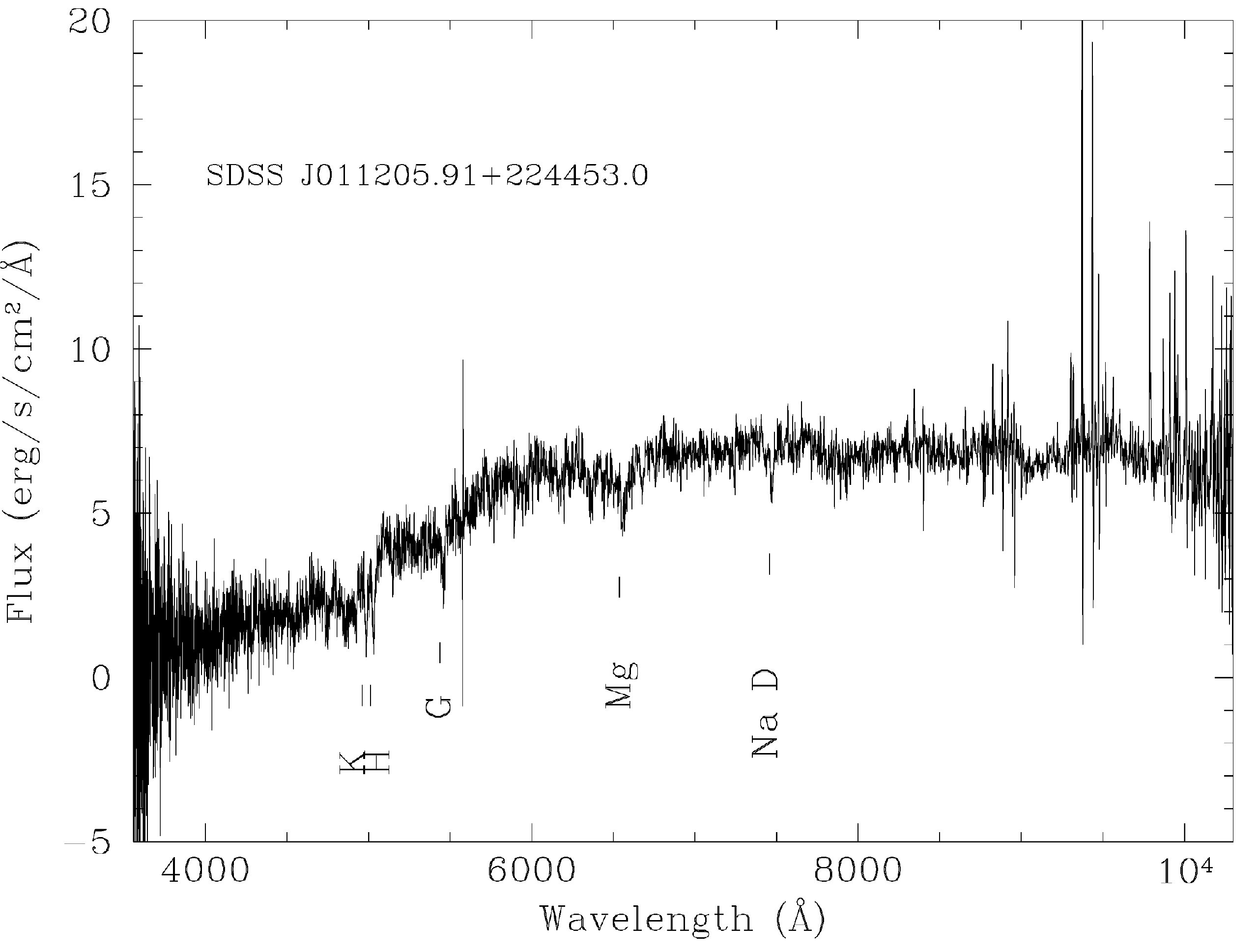}
   \caption{SDSS spectrum of the galaxy G1 (see also Fig. \ref{fig:NTT_image_R_BL0109}) at $\sim$15'' North of the BL Lac S2 0109+22. Galactic absorption lines at z=0.268 are easily detected.}
              \label{fig:G1_sdss}%
    \end{figure}

Our high quality optical spectral data confirm our previous results that no emission or absorption lines are present in the spectrum of S2 0109+22 (see Fig. \ref{fig:S0109_superimp_spectra}).
The inconsistency with the spectra presented by \cite{healey2008} and \cite{shaw2012} could be due to a target mispointing. In fact we note that at $\sim$15'' north of the target there is a r=18.3 galaxy (G1; see Fig. \ref{fig:NTT_image_R_BL0109}). The spectrum of this galaxy, taken from the Sloan Digital Sky Survey (SDSS), is very similar to that presented by \cite{healey2008} and \cite{shaw2012} and exhibits absorption lines at $z=$0.268, that is very close to $z=$0.265 alleged by these authors (Fig. \ref{fig:G1_sdss}). 

We thus conclude that the redshift z=0.265, attributed to the BL Lac object and present in NED and TeVcat database\footnote{http://tevcat.uchicago.edu/}, is wrong as it refers to the close companion galaxy. The redshift of the S2 0109+22 is therefore still unknown. Emission models of the source \citep[e.g.][]{ghisellini2011} assuming the wrong redshift should be revised.

Given the featureless spectrum of this source it becomes of interest  to set lower limits assuming that the active nucleus is hosted in a typical massive galaxy. From the spectrum obtained in 2006 (see Fig. \ref{fig:S0109_superimp_spectra}), \cite{sbarufatti2009} derived a lower limit to the redshift $z>0.25$ based on the minimum observable equivalent width ($EW_{min}$) under the assumption that the nucleus is hosted by an elliptical galaxy of M$_{R}$=-22.9 \citep{sbarufatti2006metodo}.  
Following the same procedure applied to the GTC spectrum (better combination between SNR and spectral resolution), we are now able to set a more stringent spectroscopic lower limit of $z>0.35$, based on the $EW_{min}$=0.06 $\AA$  for absorption features. 

\begin{figure}
   \centering
   \includegraphics[width=8cm]{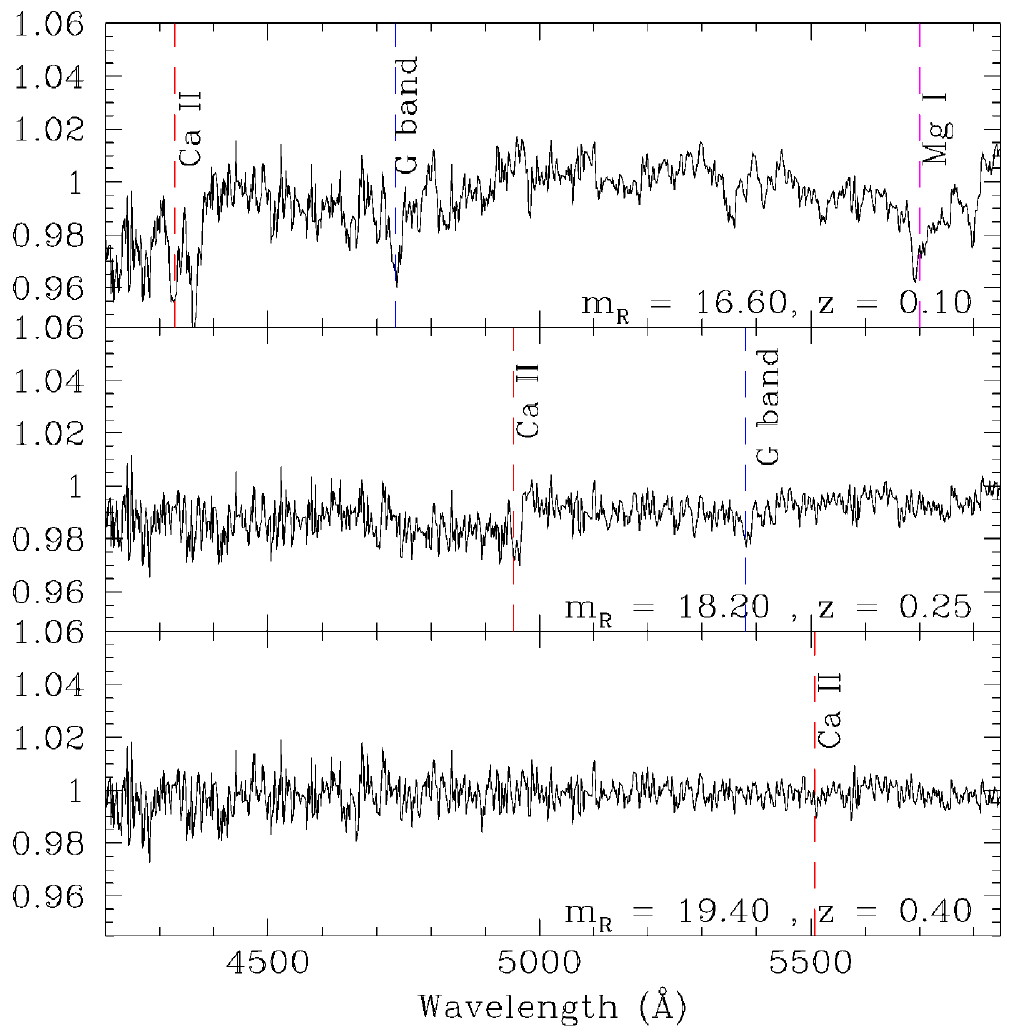}
   \caption{Simulated spectra of S2 0109+22 assuming the contribution of the host galaxy (M$_{R}$=-22.9) at three different redshifts. Apparent magnitudes, in each panels, represent the values of the host galaxy through the slit.  The main absorption features are indicated by vertical dashed lines. Ca II and G-band are still revealed for the case z=0.10 and z=0.25, but not for z=0.40.}
              \label{fig:S20109_spectrum_plus_galaxy}%
    \end{figure}

To further support this result, we report in Fig. \ref{fig:S20109_spectrum_plus_galaxy} the simulation of the superposition of the observed BL Lac spectrum with an elliptical galaxy template spectrum \citep{kinney1996} of M$_{R}$=-22.9 \citep{sbarufatti2005} at three different redshifts ($z$= 0.1, 0.25 and 0.4). The simulation takes into account the fraction of the galaxy flux lost in the slit (width= 1''). It turns out that, while for $z$=0.1 and $z$=0.25 the signature of the host galaxy is detectable, at $z$=0.4 the spectrum remains featureless. 

Finally it is worth to note that, in the environment of the galaxy G1 (see Fig. \ref{fig:NTT_image_R_BL0109}), there is a group of faint galaxies that could form a physical association at $z\sim$0.26 adopting the photometric redshifts from SDSS. Our redshift limit of S2 0109+22 is inconsistent with the redshift of this small galaxy group.
We also note that the BL Lac object lies at a projected distance of $\sim$60 kpc from G1. Spectroscopy covering region around 3500 $\AA$ could therefore reveal intervening absorption systems of Mg II (2800\AA) from the halo of G1 \citep[see e.g.][]{landoni2014, farina2014}.

Prospects of future determination of the redshift of S2 0109+22 rely on much higher SNR spectroscopy obtainable with extremely large telescopes \citep[see e.g.][]{landoni2014} or from setting stringent lower limits based on the detection of the Ly-$\alpha$ forest absorption from ultraviolet spectra \citep{danforth2010}. 

\section*{Acknowledgements}
\footnotesize{This research has made use of data obtained from the Sloan Digital Sky Survey III. Funding for SDSS-III has been provided by the Alfred P. Sloan Foundation, the Participating Institutions, the National Science Foundation, and the U.S. Department of Energy Office of Science. The SDSS-III web site is http://www.sdss3.org/. SDSS-III is managed by the Astrophysical Research Consortium for the Participating Institutions of the SDSS-III Collaboration. 
This investigation is supported by the INAF grant \''Premiale ELT 2013'' -  Rif. D.D. 33/2014.
}




\bibliographystyle{mnras}
\bibliography{0109biblio} 







\bsp	
\label{lastpage}
\end{document}